\documentclass[final,5p,times,twocolumn]{elsarticle}
\usepackage{graphicx,amssymb}
\begin{document}
\begin{frontmatter}
\title{The Production of Ultra High Energy Cosmic Rays during the Early Epochs of Radio-loud AGN} 

\author[mpp]{Hajime Takami}
\author[ccapp]{Shunsaku Horiuchi}

\address[mpp]{Max Planck Institute for Physics, F$\ddot{o}$hringer Ring 6, 80805 Munich, Germany}
\address[ccapp]{Center for Cosmology and AstroParticle Physics, The Ohio State University, 191 W. Woodruff Ave., Columbus, OH 43210, USA}

\begin{abstract}
Powerful radio-loud active galactic nuclei (AGN) with large Mpc-scale jets have been theoretically motivated as emitters of high-energy cosmic rays. Recent radio observations have established a populous class of young radio-loud galaxies with compact ($< 1$ kpc) symmetric jets that are morphologically similar to large-scale AGNs. We show that these compact AGNs, so-called compact symmetric objects (CSOs), can accelerate protons up to $10^{20}$ eV at their hot spots via a Fermi type mechanism on the assumption of efficient acceleration. The required magnetic field strengths are comparable to those derived from the minimum energy condition. We further show that the accelerated protons can escape through the photon fields of the cocoon without significant energy loss. However, the local number density of powerful CSOs is insufficient for CSOs to power the entire observed flux of ultra-high-energy cosmic rays, providing maximally only a few percent. A heavy composition of UHECRs allows more CSOs to accelerate particles to UHECR energies, but escaping the cocoon is difficult. We comment on a method that may test CSOs as UHECR sources.
\end{abstract}

\end{frontmatter}

\section{Introduction}

Radio-loud active galactic nuclei (AGN) are extremely luminous objects displaying non-thermal activity across many wavebands from radio to TeV gamma rays. The radio emission is attributed to synchrotron radiation of accelerated electrons, while the gamma-ray emission may be attributed to (i) inverse-Compton of accelerated electrons on either synchrotron photons (synchrotron self-compton, e.g., \cite{Maraschi1992ApJ397L5,Bloom1996ApJ461p657}) or external photons (external compton, e.g., \cite{Dermer1993ApJ416p458,Sikora1994ApJ421p153}), (ii) decays of neutral pions produced in hadronic interactions of accelerated protons or nuclei, e.g., \cite{Mannheim1993AA269p67,Mucke2001APh15p121} and/or (iii) synchrotron radiation of ultra-high-energy (UHE) protons, e.g., \cite{Mucke2001APh15p121,Aharonian2002MNRAS332p215,Mucke2003APh18p593}. Although leptonic models provide satisfactory fits for many AGNs, the hadronic scenario is favored for some specific ones \cite{Boettcher2009ApJ703p1168}. AGNs have therefore been extensively studied as potential sources of ultra-high-energy cosmic rays (UHECRs) \cite{Rachen:1992pg,Norman:1995aa,Berezhko2008ApJ684L69,Dermer2009NJPh11p5016,Pe'er:2009rc,Dermer2010arXiv10044249}.

Most AGN gamma-ray detections are of the blazar-class of AGNs, which are known to have relativistic jets pointed towards the observer. According to the unification model of AGNs \cite{Urry1995PASP107p803}, radio-loud Fanaroff-Riley (FR) galaxies and blazars originate from the same population, with the difference due to viewing angle. Unlike blazars, the jets in FR galaxies are not pointed towards the observer, and therefore the gamma-ray fluxes are expected to be much smaller. However, gamma rays have recently been detected from the giant lobe of a nearby FR galaxy \cite{Abdo2010Sci328p725}. \footnote{Note that the first detected non-blazar source is M87 \cite{Beilicke2004NewAR48p407,Aharonian2006Sci314p1424}, although the emission is from the core.} Considering the possibility of hadronic components in blazars, hadronic acceleration could also occur at the hotspots and/or lobes of FR galaxies. Indeed, the hot spots of strong FR II galaxies have been theoretically motivated as sources of  for some time \cite{Rachen:1992pg}. 

Interestingly, recent progress on VLBI observations have revealed the existence of compact radio-loud AGNs that are morphologically similar to FR II galaxies---e.g., they contain symmetric radio lobes and terminal hot spots---but the linear size, $LS$, is considerably smaller, $\lesssim 1$ kpc (compact symmetric objects, CSOs). Very high radio powers are observed, up to $L_{\rm syn} \sim 10^{43}$-$10^{46}$ erg s$^{-1}$, and their radio power per unit frequency has a turnover at $\sim$ GHz (GHz-peaked spectrum objects, GPS; see, e.g., Ref.~\cite{O'Dea:1998mc}). Due to these observed properties, CSOs are thought to be precursors of radio-loud AGNs, i.e., observed at an early stage of jet expansion \cite{Phillips:1982aa,Fanti:1995aa}. Flux-limited samples of radio sources find that a substantial fraction of the order of 10\% are CSOs and related compact AGNs. Since this is larger than what a simple one-to-one evolution scenario to large-scale FR II galaxies would predict, various luminosity function and evolution scenarios have been proposed \cite{Readhead1996ApJ460p612,O'Dea:1997aa}. The strong non-thermal activity, similarity to radio galaxies with large jets, and the significant fraction in flux-limited radio catalogs, motivate investigating whether CSOs can generate UHECRs. 

In this study, we discuss hadronic particle acceleration at the hot spots of CSOs. The possibility of acceleration in similar conditions has been studied by Ref.~\cite{Takahara:1990he}, where the author discussed the capability of accelerating protons up to $10^{20}$ eV at $\sim 100$ pc from the central black hole. We extend this by including the treatment of the energy source of the magnetic field, co-evolution of radio jets with cocoons, and the escape of the protons out of the cocoon. Furthermore we put this in the context of CSOs, and discuss whether CSOs are able to explain a significant fraction of the observed UHECRs based on recent data of CSOs and UHECR observatories. We consider mainly protons as the composition of UHECRs, but we also discuss applications to heavy compositions.

The paper is organized as follows. In Section 2 we review our adopted model for the hydrodynamical evolution of CSOs. In Section 3 we discuss our investigation of the generation of cosmic rays at CSO hotspots, including magnetic field energetics, cosmic ray acceleration, cosmic ray energy loss, and cosmic ray escape through the cocoon. In Section 4 we discuss whether CSOs can power the observed UHECR and discuss the impact of heavy composition. We conclude with a summary in Section 5.

\section{CSO model} \label{cso_model}
AGN jets are thought to be relativistic, $v_j \sim c$, and are decelerated by shocks produced by interactions of the jet with the interstellar medium of the host galaxy. In general, terminal shocks as well as reverse shocks (RSs) are generated, where large fractions of the jet kinetic energy are dissipated. The shocked hot plasma subsequently expands sideways to produce the cocoon. Substantial energy is also converted to accelerated particles, as shown by the observed non-thermal radiation from the shocked jet region near the RSs, so-called hot spots (see Ref.~\cite{Kino:2003gz} for a schematic view). We consider UHECR generation at the RSs which generates the hot spot. 

In order to physically describe the evolution of CSOs, we adopt the simple model worked out in Ref.~\cite{Begelman:1989aa}, originally for classical radio-loud galaxies expanding in an ambient medium. This model consists of 3 equations. The first is the balance between the momentum flux of the jet and the ram pressure of the ambient medium, 
\begin{equation}
\frac{L_j}{v_j} = n_a m_p {v_h}^2 A_h,
\label{eq:jh} 
\end{equation}
where $L_j$, $n_a$, $m_p$, $v_h$, $A_h$ are the jet kinetic power, the number density of ambient medium, the proton mass, the advance velocity of the hot spot, and the cross-sectional area of the bow shock at the end of the cocoon, respectively. The second is the balance between the pressure in the cocoon, $p$, and the ram pressure,
\begin{equation}
p = n_a m_p {v_c}^2, 
\end{equation}
where $v_c$ is the sideways expansion velocity of the cocoon (the speed of the shock driven by the overpressured cocoon). The third is energy conservation, 
\begin{equation}
p V \simeq \frac{L_{\rm tot} t}{( \hat{\gamma} - 1 )}, 
\end{equation}
where $V$, $\hat{\gamma}$, $L_{\rm tot}$, and $t$ are the cocoon's volume, the adiabatic index of the relativistic cocoon's fluid ($\hat{\gamma} = 4/3$), the total luminosity of the jet, and time, respectively. 

The $L_j$ is related to $L_{\rm tot}$ by $L_j = \eta_{\rm K} L_{\rm tot}$, where $\eta_{\rm K}$ is the fraction of the jet kinetic power to the total jet power. Eq.~\ref{eq:jh} describes the dynamics of the jet head and we assume $L_{\rm tot}$ to be constant during the active phase of CSOs. Since we focus on CSOs with $LS \lesssim$ 1 kpc inside elliptical galaxies, the density of the hot ambient medium can be assumed to be constant, $n_a \sim 0.1$ cm$^{-3}$ \cite{Mathews:2003nm}. Following these assumptions and that ${l_c}^2 \propto t$ \cite{Kawakatu:2006qc}, which was proposed to reproduce the initial phase of jet propagation according to simulations of Ref.~\cite{Scheck:2001cq}, leads to $v_h \propto t^0$ and $A_h \propto t^0$. $l_c$ is the transverse size of the cocoon. The constant $v_h$ is in good agreement with observations, which typically find $v_h \sim 0.1 c$ (see, e.g., Ref.~\cite{Owsianik:1998aa,Owsianik:1998ab}). Based on this model, the size of a particle acceleration region is estimated as 
\begin{equation}
r_h \sim \sqrt{\frac{A_h}{\pi}} = 70 \frac{ \eta_{{\rm K},0.6}^{1/2} L_{{\rm tot},46}^{1/2} }{ n_{a,-1}^{1/2} \beta_{h,-1} }~{\rm pc}, 
\end{equation}
where $\eta_{{\rm K},0.6} = \eta_{\rm K}/0.6$, 
$L_{{\rm tot},46}=L_{\rm tot}/10^{46}$ erg s$^{-1}$, 
$n_{a,-1} = n_a / 10^{-1}~{\rm cm}^{-3}$, 
and $\beta_{h,-1} = (v_h/c) / 10^{-1}$. 
Note that we assume $A_h$ to be comparable with the cross-sectional area 
of the hot spot. 

\section{Cosmic Ray Generation}

\subsection{Cosmic Ray Acceleration}
In the absence of energy-loss processes (we will discuss these in section \ref{sec:energyloss}), the maximum energy of particles accelerated at the RS can be estimated by comparing the time scale to accelerate particles and the minimum time scale in which the particles stay in the acceleration region \cite{Hillas:1985is}. The former time scale is 
\begin{equation}
\tau^\prime_{\rm acc} \sim \frac{\theta_{F} r_g^\prime }{c} = 1 \times 10^{10}  \frac{\theta_{F} \epsilon_{p,20} }{ {B^{\prime}_{-3}} \Gamma} \, {\rm s}, 
\label{tauacc}
\end{equation}
where $r_g^\prime$, $\epsilon_{p,20} = \epsilon_p / 10^{20}$ eV, $B^\prime_{-3} = B^\prime / 10^{-3}$ G, and $\Gamma$ are the Larmor radius of the particle in the RS rest frame (RSF), the proton energy in the observer frame (OF), the strength of magnetic field in the RSF, and the Lorentz factor of the RS in the OF, respectively. Throughout the paper, primes ($\prime$) explicitly denote quantities in the RSF. Since the velocity of the shocked jet in the OF is $\beta_h$ and in the RSF is $1/3$ for the strong shock limit \cite{1959flme.book.....L,Blandford:1976uq}, the velocity of the RS in the OF, $\beta_{\rm RS}$, is 
\begin{equation}
\beta_{\rm RS} = \frac{1/3 - \beta_h}{1 - (1/3) \beta_h} \simeq 0.25, 
\end{equation}
and thus $\Gamma = (1 - {\beta_{\rm RS}}^2)^{-1/2} \sim 1$. We take $\theta_F$ to be a constant, which is equivalent to assuming the diffusion coefficient for accelerated particles to be proportional to the Bohm diffusion coefficient; $\theta_F \gtrsim 10$ is a fairly conservative value, while $\theta_F \sim 1$ may be achieved for mildly relativistic shocks (see, e.g., Ref.~\cite{Rachen:1998fd}). Note that Eq.~(\ref{tauacc}) implies that the region to accelerate protons up to $10^{20}$ eV satisfies $LS \gtrsim 100$ pc from $LS/v_h \gtrsim \tau_{\rm acc}$ where $\tau_{\rm acc} = \Gamma {\tau}'_{\rm acc}$. 

There are two time scales that can limit the time for particles to stay in the acceleration region, the escape time $\tau'_{\rm esc} $, and the dynamical time $\tau'_{\rm dyn}$. The lower limit for $\tau'_{\rm esc} $ is given by the shock-crossing time in the upstream fluid, $\tau^\prime_{\rm esc} \sim r_h/\Gamma \beta' c$, while the dynamical time scale is $\tau'_{\rm dyn} \sim LS/\Gamma \beta_{\rm RS} c$, where $\beta' = 1$ is the velocity of unshocked jet assuming the strong shock limit. For our focus ($LS \gtrsim 100$ pc), $\tau'_{\rm esc} < \tau'_{\rm dyn}$ holds quite generally. Requiring $\tau'_{\rm acc} < \tau'_{\rm esc}$, we obtain 
\begin{equation}\label{Brequired}
B' > 2 \frac{\theta_F \beta' \epsilon_{p,20} }{ r_{h,1.8}} \, {\rm mG}.
\end{equation}
Thus, a few mG magnetic field is required to accelerated protons up to $10^{20}$ eV even if protons are accelerated efficiently ($\theta_F \sim 1$).

In fact, magnetic fields of a few mG have been estimated for CSO hotspots, from the minimum energy condition (e.g., \cite{Perucho:2002aa}), 
\begin{equation} \label{Bmin}
B = \left( \frac{6 \pi A L_{\rm hs}}{V_{\rm hs}} \right)^{2/7} 
= 2 \frac{{L_{\rm hs,44}}^{2/7}}{{r_{\rm h,1.8}}^{-6/7}} \, {\rm mG}, 
\end{equation}
where $L_{\rm hs,44} = L_{\rm hs} / 10^{44}$ erg s$^{-1}$, 
$V_{\rm hs} = 4 \pi {r_h}^3 / 3$, and 
$A = 3.1 \times 10^7$ in cgs units 
are the synchrotron luminosity in the hot spot, the volume of the hot spot, and a numerical factor, respectively. The factor $A$ is dependent on the spectral index of synchrotron radio emission per unit frequency above the turnover frequency, denoted by $\alpha$ \cite{Perucho:2002aa}. For our calculations we adopt the averaged value observed in CSOs, $\alpha = 0.73$ \cite{DeVries:1997aa}. The value of Eq.~(\ref{Bmin}) is comparable with the required value in Eq.~(\ref{Brequired}) for $\theta_F \sim 1$. Thus, the energy of protons can reach $10^{20}$ eV if they are accelerated efficiently.

\subsection{Hot Spot Energetics}
The magnetic field of Eq.~(\ref{Brequired}) corresponds to a luminosity of 
\begin{equation} \label{Bluminosity}
L_B = \frac{B'^2}{8 \pi} 4 \pi r_h^2 \Gamma^2 \beta' c \sim 2 \times 10^{45} {\theta_F}^2 \epsilon^2_{p,20} \, {\rm erg \, s^{-1}},
\end{equation}
where dependencies on parameters of order unity---$\beta'$ and $\Gamma$---have been removed for clarity. One sees the dependency on $L_{\rm tot}$ has cancelled; in fact, Eq.~(\ref{Bluminosity}) is a simple generic requirement of the magnetic luminosity, for a system accelerating protons to $10^{20}$ eV under stochastic acceleration mechanisms, i.e., those that must confine the accelerated particle \cite{Norman:1995aa,Waxman:2003uj,Pe'er:2009rc}. Since the magnetic field is powered by the jet energetics, one immediately sees that $L_{\rm tot} \gtrsim 10^{45.5}$ erg s$^{-1}$ for $\theta_F \sim 1$ is required to accelerate protons to $10^{20}$ eV, where $L_B = \eta_{\rm B} L_{\rm tot}$ and $\eta_{\rm B} \lesssim 1$ is a parameter. This is a large luminosity but it is not unreasonable, as we discuss next. Note also that this condition is relaxed for heavy nuclei, which have smaller Larmor radii for a given particle energy; for example, the required $L_B$ for pure iron composition would be $10^{42.4}$ erg s$^{-1}$ \cite{Pe'er:2009rc}.

The value of $L_{\rm tot}$ can be derived from the radio synchrotron luminosity, $L_{\rm syn}$. The total energy of synchrotron radiation is 
\begin{equation}\label{Lsynchrotron}
L_{\rm syn} = \frac{4 \sigma_{\rm T} U_B U_e}{ 3 m_e c} \frac{4}{3} \pi r_h^3 f(\gamma), 
\end{equation}
where $m_e$ and $\sigma_T$ are the usual electron mass and Thomson scattering cross section, $U_B$ and $U_e$ are the hot spot magnetic field and accelerated electron energy densities (both in the OF), and $\gamma$ is the Lorentz factor of accelerated electrons, respectively. Here, 
\begin{equation}
f(\gamma) = \left[ \int_{\gamma_{\rm min}}^{\gamma_{\rm max}} d\gamma \gamma \frac{dN_e}{d\gamma} (\gamma) \right]^{-1} \int_{\gamma_{\rm min}}^{\gamma_{\rm max}} d\gamma \gamma^2 \frac{dN_e}{d\gamma} (\gamma), 
\end{equation}
and $dN_e/d\gamma$ is the spectrum of accelerated electrons.

While shock acceleration predicts a single power-law spectrum, we assume a broken power-law spectrum,
\begin{eqnarray}
\frac{dN_e}{d\gamma} \propto 
\left\{
\begin{array}{ll}
\gamma^{-s_1} & \quad  1 \leq \gamma < \gamma_{\rm br}
\\
\gamma^{-s_1-1} & \quad  \gamma_{\rm br} \leq \gamma \leq \gamma_{\rm max}
\end{array} \right.
\end{eqnarray}
reflecting the expected suppression at high energies due to synchrotron energy losses (see, e.g., \cite{Stawarz2008ApJ680p911}). The value of $\gamma_{\rm br}$ is derived by equating the energy loss time, $\tau_{\rm esync} = 3 {m_e}^2 c^3 / 4 \sigma_{\rm T} \epsilon_e U_B $, to the time scale in which electrons escape from the hot spot, $t_s$. The maximum $\gamma$ of electrons is estimated by $\tau_{\rm acc} < \tau_{\rm esync}$ to be $\gamma_{\rm max} = 4 \times 10^9 {\theta_F}^{-1/2} {B_{-3}}^{-1/2}$. The synchrotron photon spectral slope $\alpha$ is related to $s_1$ by  $\alpha = (s_1-1)/2$, and would steepen above the break frequency  $\nu_{\rm br}$ corresponding to $\gamma_{\rm br}$.

Although the escape mechanism is unclear at present, the value of $t_s$ can be limited. The minimum must be $r_h/c = 2 \times 10^2 {\eta_{\rm K,0.6}}^{1/2} {L_{\rm tot,46}}^{1/2} {n_{a,-1}}^{-1/2} {\beta_{h,-1}}^{-1}$ yr, and the maximum is the age of the source, $\sim LS/v_h = 3 \times 10^3 LS_2 {\beta_{h,-1}}^{-1}$ yr, where $LS_2 = LS / 10^2$ pc.
For $t_{\rm s} = 10^2$ yr, corresponding to the minimum, $\gamma_{\rm br} = 2.4 \times 10^5 {t_{s,2}}^{-1} {B_{-3}}^{-2}$. Larger ages would result in smaller $\gamma_{\rm br}$. Observations generally agree with this simple formulation. For the handful of CSOs where the hot spots and lobes are separately resolved spectroscopically, the $\nu_{\rm br}$ in the lobes are lower (older) than those in the hot spots \cite{Murgia2003PASA20p19}. This supports the framework that electrons are accelerated at the hot spots and subsequently expand sideways to form the lobe and cocoon. Note that $\alpha$ is observed to lie in $0.5 < \alpha \lesssim 1.0$ with an average of 0.73 \cite{DeVries:1997aa}, consistent with theoretical expectations of $s_1$.

Adopting the observed average $\alpha = 0.73$ ($s_1 = 2.46$) and $\gamma_{\rm max}=4\times 10^9$, we can obtain $f(\gamma) \sim 10^3$ in the range of uncertainty of $t_s$. Now, $U_B = \Gamma^2 U'_B$ and $U_e / \Gamma^2 \simeq U'_e = \eta_e L_{\rm tot} / (4 \pi {r_h}^2 \Gamma^2 \beta_e c)$, where $\beta_e$ and $\eta_e$ are electron escape velocity normalized by $c$ and the fraction of relativistic electron luminosity to the total jet power, respectively. Although $\beta_e$ is uncertain as described above, we assume it to be the velocity of the downstream fluid at the RSF for the estimation. Substituting these into Eq.~(\ref{Lsynchrotron}), we obtain,
\begin{equation}
\frac{ L_{\rm syn} }{ L_{\rm tot} } = 5 \times 10^{-3} 
\frac{ \eta_{B,0.2} \eta_{e,0.2} {L_{{\rm tot},46}}^{1/2} n_{a,-1}^{1/2} }{ \beta' \eta_{K,0.6}^{1/2} \beta_{h,-1} \beta_{e,1/3} } \frac{f(\gamma)}{10^3},
\end{equation}
where $\eta_{e,0.2} = \eta_e / 0.2$, $\eta_{B,0.2} = \eta_B / 0.2$. Thus, despite uncertainties, $L_{\rm syn}$ is $\sim 1 \%$ of $L_{\rm tot}$ in this model. Some previous works find $L_{\rm syn}/L_{\rm tot} \sim 10 \%$ \cite{Stawarz2008ApJ680p911,deYoung1993ApJ402p95}, which would imply a jet power of $L_{\rm tot} \sim 10^{45-46}$ erg s$^{-1}$ for an observed synchrotron luminosity $L_{\rm syn} \sim 10^{44}$ erg s$^{-1}$. Our adopted value of the jet power, $L_{\rm tot}  = 10^{46}$ erg s$^{-1}$, is on the high-$L$ end of the luminosity function, as one would expect for generators of UHECRs. 

\subsection{Cosmic Ray Energy Loss} \label{sec:energyloss}
Here we discuss energy-loss processes that can inhibit the acceleration of protons up to the highest energies. These include proton synchrotron radiation and scattering with ambient photon fields. For the photon field we include synchrotron radiation of electrons that are accelerated in the same region as protons, as well as the disk, torus, and stellar emissions. As we show below, protons are not prevented from being accelerated up to $10^{20}$ eV.

The time scale for proton synchrotron energy-loss is
\begin{equation}
\tau'_{\rm psyn} = \frac{3 {m_p}^4 c^3}{4 \sigma_T {m_e}^2 \epsilon_p U'_B} \sim 2 \times 10^{13} {\epsilon'_{p,20}}^{-1} ~~~ {\rm s},
\label{tausync}
\end{equation} 
adopting the magnetic field of Eq.~(\ref{Brequired}). Since $\tau'_{\rm acc} \ll \tau'_{\rm psyn}$ we confirm proton synchrotron energy loss is not important.

We normalize the synchrotron photon field from the observed non-thermal radio emission. We expect a spectral steepening at $\nu_{\rm br} = 3 \times 10^5 {B_{-3}}^{-3} {t_{s,2}}^{-2}$ GHz \cite{Jaffe1973AA26p423,Nagai2006ApJ648p148} corresponding to $\gamma_{\rm br}$, above which the spectral index steepens from $\alpha$ to $\alpha + 1/2$. Given the observed values of $\alpha$, the radiated energy peaks at $\nu_{\rm br}$. Furthermore, synchrotron photons are self-absorbed at low frequencies, and the spectral index changes to 5/2 below the turnover frequency $\nu_{\rm sa} \sim 0.5$ GHz, estimated by equating the optical depth of synchrotron self-absorption, $\alpha_{\rm sa} r_h$, with unity, where $\alpha_{\rm sa}$ is the absorption coefficient of synchrotron self-absorption given in standard textbooks, e.g., \cite{1979rpa..book.....R}. Therefore, assuming all the photons are emitted from the hot spot, the synchrotron power per unit frequency $L_{\nu}$ is
\begin{eqnarray}
L_\nu = 
\left\{
\begin{array}{ll}
N {\nu_{\rm sa}}^{-\alpha-5/2} \nu^{5/2} &~( \nu < \nu_{\rm sa} ) \\
N \nu^{-\alpha} &~( \nu_{\rm sa} \leq \nu < \nu_{\rm br} ) \\
N {\nu_{\rm br}}^{1/2} \nu^{-\alpha - 1/2} &~( \nu_{\rm br} \leq \nu )
\end{array} \right. ,
\label{sync}
\end{eqnarray}
where 
\begin{equation}
N = \frac{L_{\rm syn}}{\left( \frac{2}{7} - \frac{1}{1 - \alpha} \right) {\nu_{\rm sa}}^{1 - \alpha} + \left( \frac{1}{1 - \alpha} + \frac{1}{\alpha - 1/2} \right){\nu_{\rm br}}^{1 - \alpha}}
\end{equation}
is the normalization. The average energy density of photons is 
\begin{equation}
\bar{U}_{\rm syn} = \frac{3 L_{\rm syn}}{4 \pi {r_h}^2 c} = 2 \times 10^{-8} \frac{ n_{a,-1} \beta_{h,-1}^2 }{ L_{{\rm syn},44} \eta_{{\rm K},0.6} L_{{\rm tot},46} } \, {\rm erg \, cm^{-3}}.
\end{equation}

We treat the photon fields due to an accretion disk, dusty torus around the AGN, and stellar emission in a similar way to Ref.~\cite{Stawarz2008ApJ680p911}. Unlike many GPS quasars where strong ultraviolet (UV) emission have been directly observed, the UV emission is expected to be largely absorbed by a dusty torus in the case of GPS galaxies. We thus assume that GPS galaxies are similar in its intrinsic UV luminosity to GPS quasars, with $L_{\rm UV} \sim 10^{45}$-$10^{47}$ erg s$^{-1}$ \cite{Koratkar1999PASP111p1}. The energy density of UV photons in the acceleration region is then 
\begin{equation}
U_{\rm UV} = \frac{L_{\rm UV}}{4 \pi LS^2 c} = 
3 \times 10^{-7} \frac{L_{\rm UV,46} }{ {LS_2} } \, {\rm erg \, cm^{-3} }. 
\end{equation}
A large fraction of the UV energy density, parametrized by $\eta_{\rm IR}$, can be absorbed and re-radiated in the infrared (IR) by a dusty torus. The energy density is
\begin{equation}
U_{\rm IR} = 9 \times 10^{-8} \frac{ \eta_{\rm IR,-0.5} L_{\rm UV,46} }{ {LS_2} } \, {\rm erg \, cm^{-3} }.
\end{equation} 
Finally, the energy density of star light photons is $\bar{U}_{\rm star} = 3 L_{\rm V} / 4 \pi {r_s}^2 c$, where $r_s$ is the core radius of the stellar distribution which correlates with the V-band luminosity of the host galaxy $L_{\rm V}$ as $L_{\rm V} / 10^{45}$ erg s$^{-1}$ $\sim r_s / 1$ kpc \cite{deRuiter2005AA439p487}. Therefore,
\begin{equation}
\bar{U}_{\rm star} = 8 \times 10^{-10} {L_{\rm V,45}} \, {\rm erg \, cm^{-3}}.
\end{equation}
From the above estimates we see that $U_{\rm UV} > U_B > U_{\rm IR} > \bar{U}_{\rm syn} >{U}_{\rm star}$.

The time scale for inverse-Compton energy loss is thus dominated by ${\tau'_{\rm ICS}}^{\rm UV} \approx 6 \times 10^{12} {\epsilon_{p,20}}^{-1} {L_{{\rm UV},46}}^{-1} {LS_2}^2$ s, which is still much larger than $\tau'_{\rm acc}$. Thus, ICS on the various photon fields can be neglected.

\begin{figure}[t]
\begin{center}
\includegraphics[clip,width=0.95\linewidth]{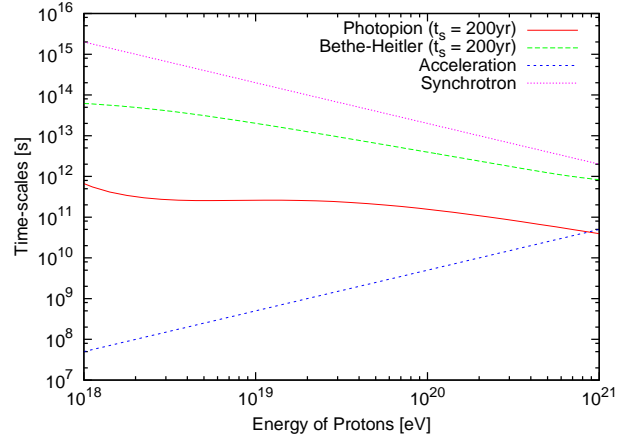}
\caption{Bethe-Heitler pair creation energy-loss time scale and mean free time of photopion production in the CSO hotspot photon field. We consider synchrotron photons, central accretion disk UV, dusty torus IR, and stellar light. The synchrotron cooling and acceleration times are also shown for comparison. The acceleration of protons is limited at $\sim 10^{20}$ eV by the time scale for particles to stay in the acceleration (see Eq. \ref{Brequired}). Here, we fix $B = 2$ mG, $r_h = 70$ pc, $L_{\rm syn} = 10^{44}$ erg s$^{-1}$, $\left< \epsilon \right>_{\rm UV} = 10$ eV, $L_{\rm UV} = 10^{46}$ erg s$^{-1}$, $\left< \epsilon \right>_{\rm IR} = 0.1$ eV, $L_{\rm IR} = \eta_{\rm IR,-0.5} L_{\rm UV} = 10^{45.5}$ erg s$^{-1}$, and $LS = 100$ pc. \label{fig}}
\end{center}
\end{figure}

In Fig.~\ref{fig} we show the energy-loss time scale of Bethe-Heitler pair creation and the mean free time of photopion production in the synchrotron photon field, the UV, and IR radiation fields. The spectra of the UV and IR radiation are assumed to be thermal distributions with the average photon energies of $\left< \epsilon \right>_{\rm UV} = 10$ and $\left< \epsilon \right>_{\rm IR} = 0.1$ eV, respectively. In terms of target photon number density, synchrotron photons play the dominant role, which in turn depend on the value of $t_{\rm s}$. To be cautious, we use the minimum $t_{\rm s}=2 \times 10^2$ yr, 
which corresponds to the direct escape of accelerated electrons.
We find that photopion production is the dominant energy loss mechanism, but it is not important up to and including $10^{20}$ eV. This is true between the minimum $t_s$ and the maximum $t_s$. 

We conclude the acceleration capability is maintained under our adopted CSO evolution scenario; that is, until the hot spot dissipates or the dynamics change. The former can arise if the AGN activity terminates \cite{Reynolds:1997yq}. The dynamics can change if $LS$ exceeds $ \sim 1$ kpc or $LS/v_h \gtrsim 10^4$ yr. This is because $A_h$ is unchanged in our CSO model and the photon fields become weaker with increasing radius. When the distance of the hot spot exceeds $\sim1$ kpc, the approximation of a constant $n_a $ is no long valid, and $A_h$ is no longer constant. This phase is not treated in this paper　
because we focus on the early epochs of radio-loud AGNs.

\subsection{Cosmic Ray Escape through the Cocoon} \label{escape}
In order to be a viable UHECR source, the protons accelerated in the hot spot must survive their escape through the cocoon. The lobe size is characterized by the radius of the cocoon, 
\begin{equation}
l_c = \left( \frac{ 8 L_{\rm tot} LS^2 }{ 3 \pi m_p n_a {v_h}^3} \right)^{1/4} \simeq 
2 \times 10^2 \frac{ L_{{\rm tot},46}^{1/4} LS_2^{1/2} }{ n_{a,-1}^{1/4} v_{h,-1}^{3/4} } \, {\rm pc},
\end{equation}
and we compare this with various relevant distance scales of the proton. 

Firstly, if the magnetic energy is conserved during the CSO evolution and cocoon formation, the energy density of magnetic field in the cocoon is estimated as 
\begin{equation}
U_{\rm B}^c = \frac{\eta_B L_{\rm tot} t}{2 \pi {l_c}^2 LS} \simeq 1 \times 10^{-6} \frac{\eta_B {n_{a,-1}}^{1/4} {L_{{\rm tot},46}}^{3/4}}{{LS_2}^{1/2} {v_{h,-1}}^{1/4}} ~~{\rm erg}~{\rm cm}^{-3}, 
\end{equation}
adopting a volume of $2 \pi {l_c}^2 LS$. Note that the shape of the cocoon is approximated to a cylinder in the CSO evolution model described in Section \ref{cso_model}. This yields a magnetic field strength of $\sim 3$ mG field for $\eta_B = 0.2$, which is unnatural given the hot spot field is $\sim 2$ mG. In reality, the magnetic energy is likely dissipated or leaking from the cocoon at some non-trivial time scale. In addition, it is difficult to realize a uniform magnetic field over this size, and a strong random or turbulent field component is expected. If the coherent length is smaller than the Larmor radius of the UHECR proton, the protons propagate in the lobe diffusively and can escape. Even in the extreme case of a uniform magnetic field, one can estimate that the Larmor radius of a UHECR proton is larger than $l_c$ for $B < 1$ mG.

The UV energy density is the largest of all energy densities (including photons and magnetic field; $U_{UV}$, $U_{IR}$, $\bar{U}_{\rm syn}^c$, $U_{B}^c$), and we find that $c \tau_{ICS}^{UV} \simeq 6 \times 10^4 {\epsilon_{p,20}}^{-1} {L_{\rm UV,46}}^{-1} {LS_2}^2$ pc. This is much larger than $l_c$, showing radiative losses are unimportant. Bethe-Heitler and photopion production energy losses can be shown to be similarly unimportant. To be conservative, we use an upper limit on the number density of the synchrotron photons as described below. Observationally, the radio emission energy from the lobe (cocoon) is weaker than that from the hot spot, i.e., ${L^c_{\rm syn}} < L_{\rm syn}$, and $\nu_{\rm br}$ is smaller in the cocoon. Thus, spectral modeling of Eq. \ref{sync} with $t_s \sim LS/v_h = 3 \times 10^3 LS_2 {\beta_{h,-1}}^{-1}$ yr gives an upper limit of the radio photons in the cocoon. Note that the change of $\nu_{\rm sa}$ is small and does not affect the normalization of the synchrotron photons. For Bethe-Heitler pair creation, the energy-loss length is at least $10^4$ pc at $10^{20}$ eV. Similarly, the mean free path of protons for photopion production in this radio field is at least $10^3$ pc at $10^{20}$ eV. Therefore, neither Bethe-Heitler nor photopion production affect the escape of $10^{20}$ eV protons.

\section{Discussion}

In this paper we focused on the early epochs of radio-loud AGNs, namely CSOs, and demonstrated that the physical conditions in their hot spots allow for the acceleration and escape of protons at $\sim 10^{20}$ eV. We now discuss the total energetics and the capability of confirming this scenario. 

The energy budget (per volume) required to reproduce the observed flux of the UHECR at Earth is nominally $\mathcal{E} (\epsilon_p > 10^{19} {\rm eV}) \sim 2 \times 10^{44}$ erg Mpc$^{-3}$ yr$^{-1}$ \cite{Waxman1995ApJ452L1,Berezinsky2006PRD74p043005,Murase:2008sa}. Uncertainties of the source spectrum introduce a factor 2 uncertainty to this value. Since UHECRs with energies above $6 \times 10^{19}$ eV travel maximally 200 Mpc because of interactions with cosmic microwave background (CMB) photons \cite{Greisen:1966jv,Zatsepin:1966jv}, whether CSOs are the main sources of the observed UHECR depends on the CSO density in the local Universe.

Since CSOs are a subclass of radio galaxies, the number density of CSOs cannot be larger than that of radio galaxies. The number density of radio galaxies with luminosities above $L_{\rm syn} \sim 10^{44}$ erg Mpc$^{-3}$ in the local Universe is $\approx 9 \times 10^{-9}$ Mpc$^{-3}$ \cite{Sadler2002MNRAS329p227}. The luminosity function of GPSs was estimated by Ref.~\cite{Snellen2000MNRAS319p445}, assuming that their cosmological evolution is similar to large-scale jet objects. Following this result, the number density of GPSs in the local Universe is estimated as $10^{-10}$-$10^{-11}$ Mpc$^{-3}$. Note that this local GPS to radio galaxy ratio is at least one order of magnitude smaller than observed in catalogs of high-redshift radio galaxies \cite{Readhead1996ApJ460p612,O'Dea:1997aa}. Now, the total energetics of UHECR from CSOs is $\mathcal{E} = {\tilde \eta}_p L_{\rm tot} n_{\rm cso}$, where ${\tilde \eta}_p$ is the fraction of the total luminosity $L_{\rm tot}$ that is given to protons with $\epsilon_p > 10^{19}$ eV. Requiring CSOs to power the observed UHECR energetics, and that ${\tilde \eta}_p < \eta_B = 0.2$, gives that $n_{\rm cso} \gtrsim 3 \times 10^{-9}$ Mpc$^{-3}$. The required number density is larger than the estimated number density of GPSs. While the uncertainties are large, adopting the local CSO number density to be $10^{-10}$ Mpc$^{-3}$, maximally only a few percent of the total UHECR flux can be due to CSOs.

\begin{figure}[t]
\begin{center}
\includegraphics[clip,width=0.95\linewidth]{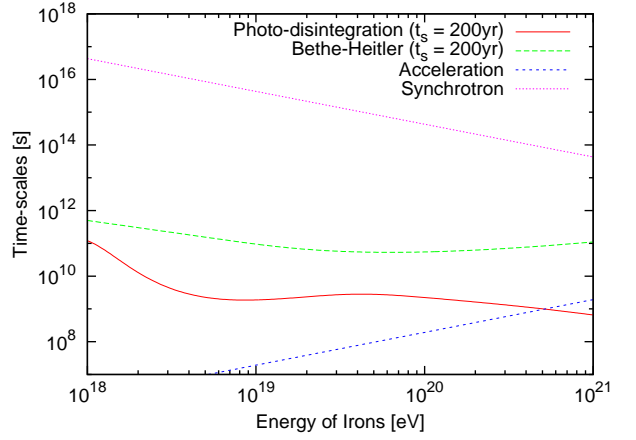}
\caption{Same as Fig. \ref{fig}, but for irons. The photopion mean free time has been replaced by the more important photo-disintegration mean free time.}
\label{fig2}
\end{center}
\end{figure}

The situation changes for heavy nuclei composition of UHECRs. A recent result of the Pierre Auger Observatory (PAO) indicates that UHECRs are dominated by heavy nuclei \cite{Abraham2010PRL104p091101} (although the High Resolution Fly's Eye (HiRes) reports a proton-dominated composition \cite{Abbasi2005ApJ622p910,Belz:2009}). Heavy nuclei dominated composition relaxes the total energy budget of UHECRs. An effect is to decrease Eq.~(\ref{Bluminosity}), and thus decrease the required CSO total luminosity $L_{\rm tot}$ \cite{Pe'er:2009rc}. The number density of CSOs above the threshold luminosity then increases, so that CSOs contribute significantly more towards the UHECR energy budget. The relative importance of CSOs depends on the luminosity function of other potential UHECR sources. Although quantitative discussions of the effects of heavy nuclei depend on the mixture of nuclei and are beyond the present paper, we give a brief discussion of the acceleration and escape of iron at the hotspot below.

Figure \ref{fig2} is similar to Figure \ref{fig} but for irons, showing the time scales for acceleration, synchrotron energy-loss, Bethe-Heitler pair creation energy-loss, and the mean free time for the photo-disintegration of iron. We find that the acceleration of iron nuclei to $10^{20}$ eV is possible. The acceleration time scale scales as $Z^{-1}$ while the synchrotron time scale scales as $A^4 Z^{-2}$ (see Equations \ref{tauacc} and \ref{tausync}) which work in favor for iron nuclei. In order to calculate the mean free time for photo-disintegration, we adopt the parametrized cross-section in Ref.~\cite{Rachen1996phd}. Although the cross-section for iron at the first peak of the GDR is $\sim 100$ times larger than the cross-section for protons at the delta resonance, the acceleration time scale is sufficiently small for iron, and iron can be accelerated to $10^{20}$ eV. 

\begin{figure}[t]
\begin{center}
\includegraphics[clip,width=0.95\linewidth]{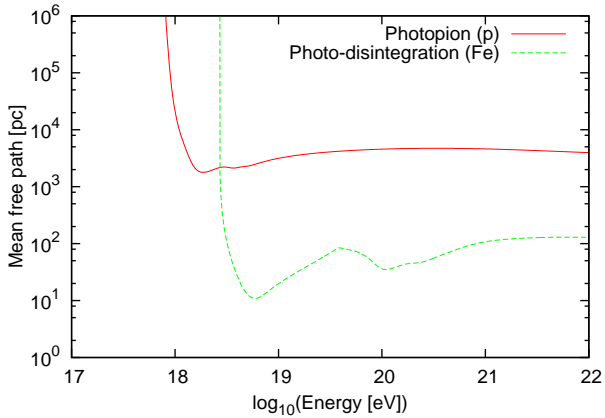}
\caption{Mean free paths of photopion production of protons 
and photo-disintegration of irons in the IR photon field 
which is approximated as a delta function with $\epsilon_{\rm IR} = 0.1$ eV.}
\label{fig3}
\end{center}
\end{figure}

However, the $10^{20}$ eV irons cannot escape from the cocoon. We demonstrate this using IR photons. Approximating the IR photon spectrum as a delta function at $\epsilon = 0.1$ eV, we calculate the mean free path of irons for photo-disintegration. This is shown in Figure \ref{fig3}. The mean free path of protons for photopion production is also shown for comparison. The mean free path of irons in the IR photon field is minimized at $\sim 10^{19}$ eV due to GDR to $\approx 10$ pc. Also, this field predicts that the mean free path of $10^{20}$ eV iron nuclei is less than $10^2$ pc due to baryonic resonance. This simple investigation shows that iron nuclei would photo-disintegrate before escaping from the cocoon, of length $l_c \simeq 2 \times 10^2$ pc. Since photo-disintegration is the process of nuclear breaking, it is possible that nuclei with lower nuclear number and lower energy could diffusively escape from the cocoon. We conclude that while CSOs could be high-energy nuclei emitters, the details depend largely on the structure of magnetic field in the cocoon.  

The recent PAO data indicates small deflections of UHECRs by Galactic magnetic field and intergalactic magnetic field (IGMF) \cite{Cronin:2007zz,Abraham:2007si,Kashti:2008bw,George:2008zd,Ghisellini:2008gb,Takami:2008ri,Abreu2010arXiv10091855} (but see Refs. \cite{Kotera:2008ae,Ryu:2009} for the possibility of spurious correlation). Thus, the correlation between the arrival directions of UHECRs and the position of CSO candidates could be a powerful way to confirm our scenario. Since CSOs are subdominant sources of UHECRs, much more detected UHECRs are required. Furthermore, the intermittency of CSO activity with the duration of $10^4$-$10^5$ yr has been suggested to explain the observed overabundance of CSOs compared to predictions of a uniform evolution scenario to large-scale radio sources \cite{Reynolds:1997yq}. If the active phase is shorter than the time-delay of UHECRs during propagation, CSOs which emitted UHECRs are possibly inactive when the UHECRs arrive at Earth. Note that there is large uncertainty on IGMF and the time-delay of UHECRs strongly depends on IGMF modellings \cite{Sigl:2003ay,Dolag:2004kp,Takami:2005ij,Ryu:2008hi,Kotera:2008ae}. In either case, we may not observe active CSOs, but instead CSO remnants \cite{Kunert-Bajraszewska2005AA440p93}, towards the directions of UHECRs. 

\section{Summary}

In summary, we investigated CSOs as a possible source of UHECRs, demonstrating that their hot spots may accelerate protons to $10^{20}$ eV. Although the local number density of CSOs capable of accelerating UHECRs contains larger uncertainties, it is generally insufficient to power the entire UHECR observed on Earth. An iron composition could increase the contribution from CSOs to the UHECR flux, but their escape through the CSO cocoon is made difficult by photo-disintegration. Future investigations of the CSO luminosity function, correlation with faint CSO remnants, and $\gamma$-ray produced simultaneously with UHECRs may test the CSOs as UHECR sources.

\subsubsection*{Acknowledgements:}
We are grateful to J.~F.~Beacom, N.~Kawakatu, and K.~Murase for useful discussions and comments. H.T.~and S.H.~thank the hospitality of the Institute for the Physics and Mathematics of the Universe (IPMU) where initial parts of this work took place. S.H.~is partially supported by NSF CAREER Grant PHY-0547102 (through JFB).

\bibliographystyle{model1a-num-names}
\bibliography{ms.bib}
\end{document}